# Fast determination of thickness through scanning moiré fringe in scanning transmission electron microscopy

Pengfei Nan, Zhiyao Liang, Yue Zhang, Yangrui Liu, Dongsheng Song, Binghui Ge[*]

Information Materials and Intelligent Sensing Laboratory of Anhui Province, Key Laboratory of Structure and Functional Regulation of Hybrid Materials of Ministry of Education, Institutes of Physical Science and Information Technology, Anhui University, Hefei, 230601, China.

(*Electronic mail: bhge@ahu.edu.cn)

## Abstract

Sample thickness is an important parameter in transmission electron microscopy (TEM) imaging, for interpreting image contrast and understanding the relationship between properties and microstructure. In this study, we introduce a method for determining thickness in scanning transmission electron microscopy (STEM) mode based on scanning moiré fringe (SMF). Sample thickness can be determined in situ in the medium magnification using focal-series SMF imaging, with beam damage and contamination avoided to a large extent. This method provides a fast and convenient way for determining thickness in TEM imaging, which is particularly useful for beam-sensitive materials such as Metal-Organic Frameworks.

**Keywords**：Thickness determination, Scanning moiré fringe, Beam-sensitive materials, Scanning transmission electron microscopy

**Introduction**

Transmission electron microscopy (TEM) is a powerful method in modern science for characterizing the microstructure of materials. The introduction of aberration correctors at the end of the last century brings us into a new era (Haider, Rose, Uhlemann, Kabius, *et al.*, 1998, Haider, Rose, Uhlemann, Schwan, *et al.*, 1998, Uhlemann & Haider, 1998, Haider *et al.*, 2000). Several imaging methods, particularly negative Cs imaging (Jia *et al.*, 2003, Jia & Urban, 2004), annular dark-field imaging (Ishikawa *et al.*, 2011), and integrated differential phase-contrast imaging (Lazić *et al.*, 2016, Close *et al.*, 2015), enrich further structural characterization, thereby aiding the observation of light atoms, such as oxygen, lithium, and hydrogen, which is unfeasible with conventional TEM due to the weak scattering capacity for the low atomic number atoms. Recently, electron ptychography imaging with a pixel-array detector achieved a resolution of 0.39 Å (Jiang *et al.*, 2018). Additionally, with the help of a direct electron detection camera, cryo TEM has become a popular method for resolving the structure of biomolecules for structural biology (Li *et al.*, 2013, Liao *et al.*, 2013).

Although great progress has been made in electron microscopy, one major obstacle to structural reconstruction is the thickness in the practical sample due to multiple electron scattering or nonlinear imaging between the interactions of different diffraction beams (Li *et al.*, 2020), which makes direct interpretation of the experimental results difficult. This is a common question for structure determination, especially in TEM using phase-contrast imaging. Sample thickness is critical not only for interpreting TEM image contrast but also for understanding intrinsic properties of nanosized samples, such as ferroelectricity (Junquera & Ghosez, 2003) and mechanical properties (Cao *et al.*, 2019). For example, Stachiotti et al. reported that freestanding $BaTiO_3$ thin films lost ferroelectricity below a critical thickness of approximately 4 nm, which was confirmed via TEM imaging and image simulation (Li *et al.*, 2015).

To determine the sample thickness, different methods were developed. A series of simulated images were calculated within a range of thickness to fit the experimental results (Li *et al.*, 2015). However, because of the multiple effects of other imaging parameters, such as defocus, tilt, and astigmatism, this method is time-consuming and not preferred. A convergent-beam diffraction pattern (CBED) analysis can achieve 5% accuracy (Castro-fernandez *et al.*, 1985), and position averaged convergent-beam diffraction patterns are now reported to be independent of the lens aberration and effective source size (LeBeau *et al.*, 2010), but is still time-consuming and works only for crystalline specimens. Presently, electron energy loss spectroscopy (EELS) is commonly used in situ method for thickness determination using log-ratios method or K-K sum rule to obtain the relative and absolute thickness (Egerton, 2011, Egerton & Cheng, 1987), but it requires prior knowledge about the sample composition, and especially an additional and expensive attachment GIF (Gatan Image Filter) system.

In this study, a simple method for determining the sample thickness using scanning moiré fringe (SMF) in the scanning transmission electron microscopy (STEM) mode is proposed. Focal-series STEM imaging at medium magnification, such as 1 M (million), was performed, rather than at magnification as high as more than 10 M to obtain the atomic resolution, so that the sample thickness can be in situ determined with beam damage and contamination avoided to a large extent.

**Materials and Methods**

Specimens for TEM analysis are prepared by traditional mechanical polishing, dimpling, and argon ion milling with a liquid nitrogen stage. The atomic resolution high angle annular dark-field (HAADF) images are carried out at 50-200 mrad, and EELS was collected with GIF system of Gatan continuum 1065. All experiments are performed by FEI Titan Themis Z microscope equipped with probe and image correctors operated at 300 kV.

**Results and discussion**

Fig. 1a shows a model of GeSe with a trigonal system, and the atomic-resolution HAADF image shown in Fig.

1b accurately reflects the projected structure along the [001] zone axis. Because GeSe is a ferroelectric material, it has a ferroelectric domain (Fig. 1c). Note that when we change the focus, fringes can be observed (Fig. 1d). This is an SMF (Su & Zhu, 2010), which is generated by the spatial interference between the scanning grating of the electron beam and the atomic lattice of the specimen (diagram of SMF (Fig. 1e)).

SMF can be one-dimensional (Fig. 1d) or two-dimensional, and even an artificial dislocation core can be obtained with Burgers circuit similar to the one obtained by atomic-resolution image (Su & Zhu, 2010). In contrast to atomic-resolution imaging, which is usually performed at a magnification greater than 10 M, SMFs are observed at the magnification of approximately 1 M. Thus, its main advantage is that it is free of beam damage, and can also weaken contamination.

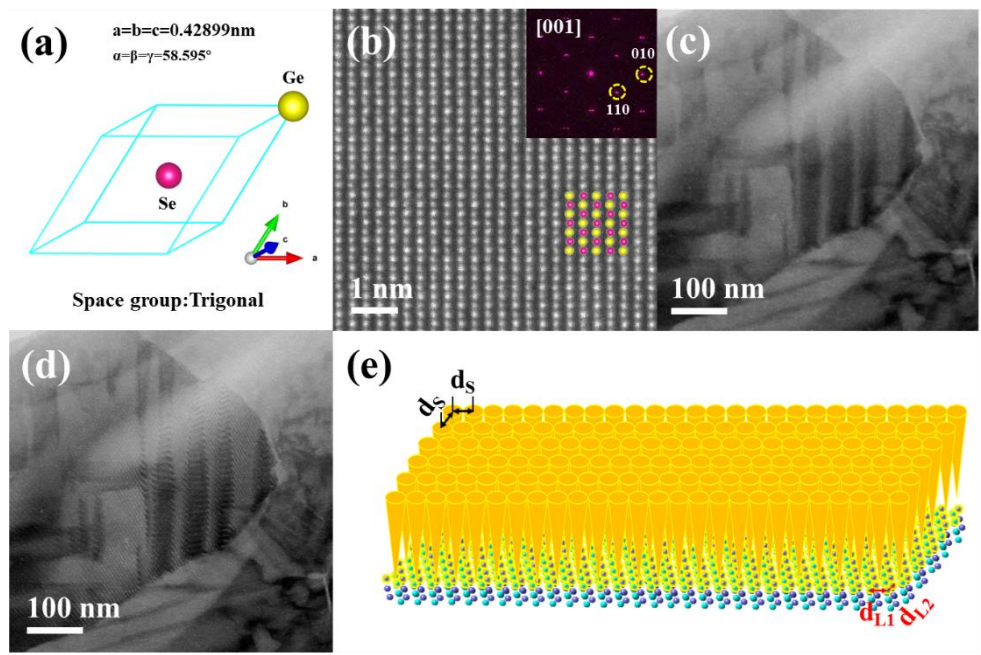

Fig. 1. (a) Structural model of GeSe, (b) atomic resolution STEM image along the zone axis [001] with corresponding diffraction pattern inset in the top right, (c) and (d) medium-magnification HAADF image and corresponding SMF, (e) schematic diagram of the SMF, where the orange cone array in the upper part represents the scanning grating, the atomic lattice is in the lower part, and the atoms at topmost layer are circled in yellow.

As shown in Figs. 1 c and d, changing the focus causes SMF to appear or disappear. Similarly, SMF can move when the focus changes (Fig. 2). When the focus value is 20 nm, as shown in the TEM monitor, there is no SMF in the area denoted by a blue rectangle, and SMFs appear at the top. When the focus is set to 30 nm, SMFs appear nearly in the entire image, including the rectangular area, but the contrast of SMF is higher in the top part than in the bottom. When the focus is set to 45 and 60 nm, the SMF continues to move down. Additionally, when the focus value changed to 70 nm, SMF almost disappears in the bottom view.

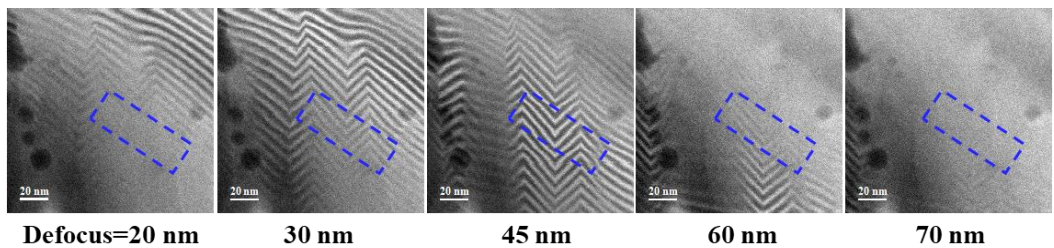

Fig. 2. Focal series of HAADF images, SMF moving from top to the bottom.

How to understand this phenomenon? We must consider the SMF forming mechanics. SMF can only be obtained when scanning raster interferes with the lattice. Here, electrons should be focused within the sample, and then the probe should interact with the sample. Specifically, a probe in the STEM mode is formed by a convergent beam (Fig. 3a). Because most electrons are distributed at the intersection indicated by the orange ellipse, this ellipse can be regarded as a probe. When the probe is overfocused, as shown in the first image in Fig. 3b, there is no strong interaction between the electron and sample, so no signal is collected by the STEM detector and no SMF is produced. When the probe enters the sample by changing the focus value, SMF appears, and when the entire probe is completely located in the sample, the contrast of SMF will be the strongest. Similarly, as the probe gradually leaves the sample, the contrast of SMF weakens and eventually disappears.

As shown in Fig. 3b, this focal series of techniques can be used to determine the sample thickness. Furthermore, in STEM imaging, the probe not only has a lateral size that limits the conventional resolution of the STEM but also has a longitudinal size that limits the depth resolution of the STEM, as indicated by the long axis of the ellipse in Fig. 3. According to Ref (Nellist *et al.*, 2008), the length of the long axis, i.e., the full-width at half-maximum of the probe along the beam direction, can be expressed as the formula $\delta_z = \frac{1.77\lambda}{\alpha^2}$ (by using different criterion, some researchers (Borisevich *et al.*, 2006) may give different formula), so in this case, semi-convergent angle $\alpha$, 25 mrad, and wavelength $\lambda$ for 300 kV, approximately 0.0197 Å, the vertical resolution is approximately 5.6 nm. The thickness should be close to 44.4 nm (i.e., 50−5.6)) (Fig. 3b). Nevertheless, the SMF method is similar to CBED, and only the crystalline information can be obtained. Considering that the total thickness of the amorphous layer is 5–10 nm, the total thickness of the sample should be 54.4−64.4 nm.

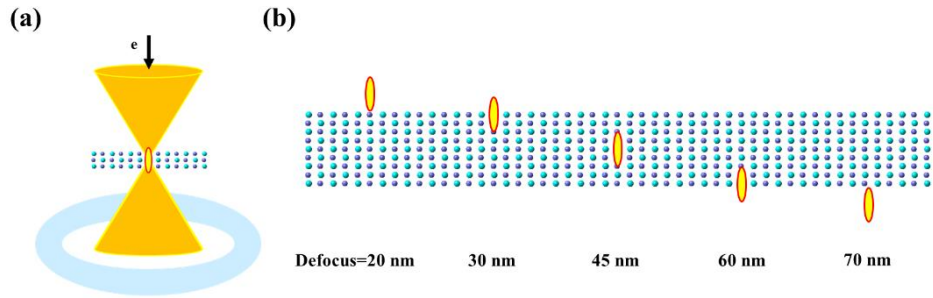

Fig. 3. (a) Schematic diagram of the probe in STEM, (b) schematic diagram of the position of the probe with respect to the sample.

To verify this thickness determination method, the EELS technique (Fig. 4a), is used to obtain the absolute thickness using the Log-ratio method (Malis *et al.*, 1988). In our experiment, the beam energy is 300 kV, the semi-convergent angle is 25 mrad, and the collection semi-angle is 33.44 mrad (Entrance Aperture: 5 mm), $Z_{eff} = \frac{\sum_i f_i z_i^{1.3}}{\sum_i f_i z_i^{0.3}} \approx 33$. The above parameters can be used to calculate the inelastic mean free path $\lambda'$, thus the absolute thickness t obtained from EELS (Fig. 4(a), $\frac{t}{\lambda'} = 0.67$) is 63 nm, which is consistent with the SMF results.

To further validate this method, a wedge-shaped Si sample is used. At first, an absolute thickness map was obtained using the GIF system (Fig. 4b). The sample thickness gradually increases from left to right, with the absolute thickness for the edge area being ~40 nm and the thickest area is ~52 nm, indicated by the white line.

For comparison, we also performed focal-series imaging in the corresponding area, and the thickness determined using the SMF method is plotted in Fig. 4c. The results of EELS agree with that of the SMF method, indicating the method's validity.

Focal-series STEM imaging has been proposed by Beyer in 2015 using atomic-resolution images to determine the sample thickness (Beyer *et al.*, 2015). In this study, we developed the SMF technique to determine the local thickness. Because SMFs always images in medium magnification, beam damage and contamination (common disadvantages of STEM) can be avoided, which is particularly useful for beam-sensitive materials, such as Metal-Organic Frameworks, lithium batteries. Furthermore, compared to CBED or EELS, the proposed method does not require mechanical tilting and specialized hardware or software, thus it can be used as an in situ mode, such as EELS.

The proposed method's accuracy is assumed to be related to two factors: the amorphous layer and the depth resolution of the STEM. The latter is limited by the semi-convergent angle of the beam and the electron wavelength, so it is expected to have an atomic vertical resolution in the future, which would enable thickness determination of crystals more accurately.

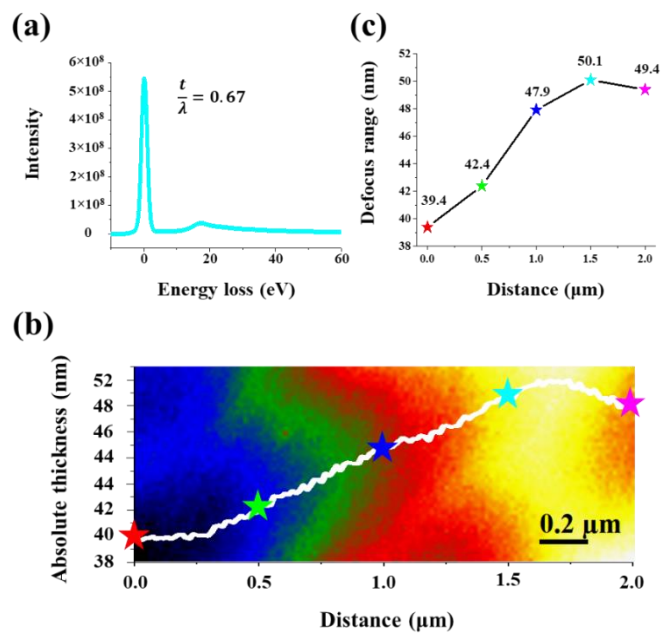

Fig. 4. (a) EELS, (b) Thickness map of wedge-shape Si sample obtained through EELS, (c) thickness for different position in (b). Thickness in (c) is obtained considering depth resolution 5.6 nm, thickness of the amorphous layer 10 nm.

## Conclusion

In summary, a method is proposed for determining sample thickness using SMF imaging. STEM conventional disadvantages, such as beam damage, and contamination, can be avoided for SMF is imaging at medium magnification. Furthermore, it only requires changing the focus, making it an in situ method that does not require additional hardware or software. Because it is fast and convenient, it is assumed that this method can be used widely, especially for beam-sensitive materials.

## Conflicts of interest

There are no conflicts to declare.

## Acknowledgements

This project is supported by the National Natural Science Foundation of China (No. 11874394), The University Synergy Innovation Program of Anhui Province (No. GXXT-2020-003). Authors thank Enago (www.enago.cn) for the English polish, and also thank Mr. Xiang Li and Prof Lipeng Hu for providing the sample.